\pdfoutput=1

\documentclass[%
    aps                ,   
    pra                ,   
    reprint            ,   
    showpacs           ,   
    superscriptaddress ,   
    nofootinbib            
]{revtex4-1}
\usepackage[T1]{fontenc}
\usepackage[utf8]{inputenc}
\usepackage[english]{babel}
\newcommand{\nhphantom}[1]{\sbox0{#1}\hspace{-\the\wd0}} 

\usepackage{amsmath}
\usepackage{amssymb}
\usepackage{braket}
\usepackage{accents}
\renewcommand{\vec}[1]{\boldsymbol{\mathbf{#1}}}    
\newcommand  {\mat}[1]{\boldsymbol{\mathbf{#1}}}    
\DeclareMathOperator{\Tr}{Tr}
\DeclareMathOperator{\Imag}{Im}

\newlength{\dhatheight}
\DeclareRobustCommand{\doublehat}[1]{%
    \settoheight{\dhatheight}{\ensuremath{\hat{#1}}}%
    \addtolength{\dhatheight}{-0.35ex}%
    \hat{\vphantom{\rule{1pt}{\dhatheight}}%
    \smash{\hat{#1}}}}

\usepackage[hyperref]{xcolor}
\usepackage{graphicx}
\usepackage[export]{adjustbox}
\graphicspath{{figures/PNG/}{figures/PDF/}{figures/}}

\usepackage[
    unicode           = true  ,
    plainpages        = false , 
    pdfpagelabels     = true  , 
    bookmarks         = true  ,
    bookmarksnumbered = true  ,
    bookmarksopen     = true  ,
    breaklinks        = true  ,
    backref           = false ,
    colorlinks        = true  ,
    linkcolor         = blue  ,
    urlcolor          = blue  ,
    citecolor         = red   ,
    anchorcolor       = green ,
    hyperindex        = true  ,
    linktocpage       = true  ,
    hyperfigures      = true
]{hyperref}
\hypersetup{
    linkcolor         = [RGB]{000 114 189} ,  
    urlcolor          = [RGB]{000 114 189} ,  
    filecolor         = [RGB]{000 114 189} ,  
    citecolor         = [RGB]{217 083 025} ,  
    anchorcolor       = [RGB]{119 171 048} ,  
    pdftitle={Steady-State Quantum Zeno Effect of Driven-Dissipative Bosons with Dynamical Mean-Field Theory},
    pdfauthor={Matteo Seclì <matteo.secli@sissa.it>}
}

\begin{document}

\title{Steady-State Quantum Zeno Effect of Driven-Dissipative Bosons with Dynamical Mean-Field Theory}

\author{Matteo Seclì}
\thanks{Now at: Department of Electrical Engineering and Computer Sciences, University of California, Berkeley, California 94720, USA}
\email[]{matteo.secli@berkeley.edu}
\affiliation{International School for Advanced Studies (SISSA), Via Bonomea 265, I-34136 Trieste, Italy}

\author{Massimo Capone}
\affiliation{International School for Advanced Studies (SISSA), Via Bonomea 265, I-34136 Trieste, Italy}
\affiliation{CNR-IOM Democritos, Via Bonomea 265, I-34136 Trieste, Italy}

\author{Marco Schirò}
\thanks{On Leave from: Institut de Physique Théorique, Université Paris Saclay, CNRS, CEA, F-91191 Gif-sur-Yvette, France}
\affiliation{JEIP, USR 3573 CNRS, Collége de France, PSL Research University, 11 Place Marcelin Berthelot, 75321 Paris Cedex 05, France}

\date{May 30, 2022}

\begin{abstract}
We study a driven-dissipative Bose--Hubbard model in presence of two-particle losses and an incoherent single-particle drive on each lattice site, leading to a finite-density stationary state. Using dynamical mean-field theory (DMFT) and an impurity solver based on exact diagonalization of the associated Lindbladian, we investigate the regime of strong two-particle losses. Here, a stationary-state quantum Zeno effect emerges, as can be seen in the on-site occupation and spectral function. We show that DMFT captures this effect through its self-consistent bath. We show that, in the deep Zeno regime, the bath structure simplifies, with the occupation of all bath sites except one becoming exponentially suppressed. As a result, an effective dissipative hard-core Bose--Hubbard dimer model emerges, where the auxiliary bath site has single-particle dissipation controlled by the Zeno dissipative scale.
\end{abstract}

\maketitle

\section{Introduction}

A variety of experimental platforms at the interface between atomic physics, quantum optics and solid-state can be theoretically described as open Markovian many-body quantum systems, where coherent quantum dynamics competes with dissipation arising from coupling to external environments~\cite{Carusotto2013,Ritsch2013,Schmidt2013,LeHur2016,Noh2017,Hartmann2016,Sieberer2016,Alyatkin2021}. In this context, the density matrix of the system evolves according to a many-body Lindblad master equation~\cite{Breuer2007}, where dissipative processes are described by a set of jump operators. For bosonic or fermionic quantum particles, these can model both single-particle processes such as pump and losses, as well as correlated effects, such as heating due to stimulated emission~\cite{Pichler2013,Poletti2013,Bouganne2020}, or multi-particle losses~\cite{Garcia-Ripoll2009,Daley2009,Kantian2009}. The latter, in particular, have been implemented both with ultracold atoms in optical lattices as well as with superconducting circuits~\cite{Lescanne2020}. In the first case, this has led to the observation of the celebrated quantum Zeno effect (QZE)~\cite{Misra1977,Beige2000}, where the effective dissipation decreases as the bare dissipation is increased~\cite{Syassen2008,Garcia-Ripoll2009,Yan2013,Zhu2014,Tomita2017,Tomita2019}. In recent years, the interest around these many-body versions of the QZE has grown~\cite{Froml2019,Rossini2021,Rosso2020,Biella2021,Muller2021,Maimbourg2021,Wasak2021,Krapivsky2019,Scarlatella2020,Rosso2021}.

From a theoretical perspective, the study of open quantum many-body systems is particularly challenging both numerically, due to the severe scaling of the Hilbert space and the need of dealing with density matrices, as well as analytically, where the combination of interactions and nonequilibrium effects limits the applicability of standard tools of many-body physics. As such, several techniques have been recently developed to solve these problems~\cite{Finazzi2015,Weimer2015,Schiro2016,Jin2016,Landa2020,Weimer2021}. A powerful approach to equilibrium and out-of-equilibrium correlated quantum systems is the dynamical mean-field theory~\cite{Georges1996,Byczuk2008,Anders2010} (DMFT), which maps the lattice many-body problem onto a quantum impurity model subject to a self-consistent dynamical mean field. Dynamical mean-field theory has been applied to a variety of nonequilibrium problems~\cite{Arrigoni2013,Aoki2014}, and it has been recently extended to open bosonic quantum systems described by a Lindblad master equation by employing a non-crossing approximation (NCA) scheme for the solution of the associated impurity problem~\cite{Scarlatella2020}. In order to expand the scope of DMFT applications to Lindblad problems, it is therefore important to develop and enlarge the set of methods that can be used to solve impurity models in presence of both Markovian and non-Markovian baths.

With this motivation, in this work we develop a Lindblad exact diagonalization (ED) impurity solver for DMFT studies of open quantum many-body systems. As we are going to discuss more in detail, this method, when compared to NCA, introduces a discretization of the DMFT self-consistent bath into a finite number of sites while treating the system--bath coupling exactly, rather than perturbatively as in NCA. ED impurity solvers have a long and successful tradition in equilibrium DMFT studies~\cite{Georges1996} and have also been used in nonequilibrium applications, for example in transport problems~\cite{Arrigoni2013,Dorda2014}.

As an application of our DMFT/ED approach, we study a driven-dissipative Bose--Hubbard (BH) model in presence of strongly correlated losses, relevant for the quantum Zeno physics. Driven and dissipative generalizations of the Bose--Hubbard model have received large attention in recent years, in particular for what concerns the coherently driven setting~\cite{LeBoite2013}, the role of incoherent driving~\cite{Biella2017,Scarlatella2019} and non-Markovian dissipation~\cite{Lebreuilly2017a}, or the role of two-particle losses leading to a decay towards the zero-density limit~\cite{Garcia-Ripoll2009,Rossini2021}. Here, we focus on the finite-density stationary state obtained by balancing the two-particle losses with a single-particle incoherent drive, as done in Ref.~\cite{Scarlatella2020}.

We show that our DMFT/ED approach recovers the qualitative features of the QZE as captured by the NCA impurity solver, and provides new insights on the problem that could not be obtained by other methods. In particular, we show that key signatures of the QZE also appear in the self-consistent DMFT bath, e.g.\@ in its effective dissipation. Furthermore, we show that, deep into the Zeno regime, the auxiliary bosonic Anderson impurity model reduces to a much simpler BH dimer problem~\cite{Secli2021}, where one of the two sites represents the rest of the system.

This manuscript is organized as follows. In Sec.~\ref{sec:model}, we introduce the driven-dissipative BH model. In Sec.~\ref{sec:dmft}, we discuss the DMFT approach to Markovian quantum many-body systems and we present an impurity solver based on exact diagonalization of the associated Linbdlad problem. In Sec.~\ref{sec:zeno}, we present our DMFT results for the driven-dissipative BH model with strong two-particle losses, after a brief review of the quantum Zeno regime of this model in Sec.~\ref{sec:hard_core_bosons}. Finally, Sec.~\ref{sec:conclusions} is devoted to conclusions.

\section{Driven-Dissipative Bose--Hubbard Model}
\label{sec:model}

We consider a lattice model of bosonic interacting quantum particles, described by the Bose--Hubbard (BH) model, coupled to local Markovian environments which induce dissipative incoherent processes on the system. As mentioned in the introduction, this model can be relevant both for ultracold atomic gases, where particles describe bosonic atoms in optical lattices, as well as for arrays of superconducting circuits, where the basic degrees of freedoms are microwave photonic excitations. In the following, we will refer to them generically as bosonic degrees of freedom.

The full evolution of the system density matrix is described by a many-body Lindblad master equation of the form
\begin{equation}
    \frac{d}{dt}\hat{\rho}^{\mathrm{BH}} =  
    -i\left[\hat{H}^{\mathrm{BH}},\hat{\rho}^{\mathrm{BH}}\right] + \doublehat{\mathcal{L}}_D^{\mathrm{BH}}\hat{\rho}^{\mathrm{BH}}
    \label{eq:lindblad_equation}
\end{equation}
where the first term describes the Hamiltonian evolution while the second one accounts for drive and dissipation. For a BH model, the Hamiltonian can be written as a local term diagonal in the occupation basis $\hat{n}_{\vec{r}} = \hat{a}_{\vec{r}}^{\dagger}\hat{a}_{\vec{r}}$, plus a tunneling contribution:
\begin{equation}
    \hat{H}^{\mathrm{BH}}
    = \sum_{\vec{r}} \hat{H}_{\mathrm{loc}}(\hat{n}_{\vec{r}})
    - \frac{J}{z} \sum_{\braket{\vec{r},\,\vec{r}'}}
    \hat{a}_{\vec{r}}^{\dagger}\hat{a}_{\vec{r}'}
    \label{eq:bh_model}
\end{equation}
where $\hat{H}_{\mathrm{loc}}(\hat{n}_{\vec{r}})=\omega_0\hat{n}_{\vec{r}}+ U\hat{n}_{\vec{r}}^2$, $\omega_0$ is a local energy term (the chemical potential for the atoms or the cavity frequency in the cQED context), $U$ is the strength of the interaction, $z$ is the number of nearest-neighbors and $J/z$ is the tunneling amplitude among any two of the nearest-neighbors indicated by $\braket{\vec{r},\,\vec{r}'}$ in the sum. 

In this work, we assume the dissipation is written as a sum of local terms, i.e.
\begin{align}
    \doublehat{\mathcal{L}}_D^{\mathrm{BH}}\hat{\rho}^{\mathrm{BH}} = 
    \sum_{\alpha\vec{r}} \left(
        \hat{L}_{\alpha\vec{r}}\hat{\rho}^{\mathrm{BH}}\hat{L}^{\dagger}_{\alpha\vec{r}} -
        \frac{1}{2}\left\lbrace \hat{L}^{\dagger}_{\alpha\vec{r}}\hat{L}_{\alpha\vec{r}},\hat{\rho}^{\mathrm{BH}} \right\rbrace
    \right)
    \label{eq:lindblad_term_dissipator}
\end{align}
where we allow for two (labelled by $\alpha=1,2$) different dissipative processes describing respectively a single-particle incoherent drive with amplitude $P_1$, giving rise to a jump operator of the form
\begin{align}
    \hat{L}_{1\vec{r}}=\sqrt{P_1}\hat{a}^{\dagger}_{\vec{r}} \,,
    \label{eq:jump_lattice_pump}
\end{align}
and two-particle losses (or gain saturation in the semiclassical limit) with amplitude $\Gamma_2$, leading to a jump operator of the form
\begin{align}
    \hat{L}_{2\vec{r}}=\sqrt{\Gamma_2}\hat{a}_{\vec{r}}\hat{a}_{\vec{r}} \,.
    \label{eq:jump_lattice_loss}
\end{align}
The resulting driven-dissipative BH model was studied in absence of drive in Refs.~\cite{Garcia-Ripoll2009,Rossini2021}, while the driven case, leading to a finite-density stationary state, was considered in Ref~\cite{Scarlatella2020}. The latter is the regime that we will focus on in this work. In the next Section we introduce the method we use in our investigation, and we discuss the main results in Sec.~\ref{sec:zeno}.

\section{Dynamical Mean-Field Theory for Markovian Bosons}
\label{sec:dmft}

In this Section, we describe the DMFT approach to driven-dissipative bosons, including the mapping onto a Markovian quantum impurity model and the exact-diagonalization impurity solver that we developed to tackle this problem. Though we specifically refer to the model introduced in Sec.~\ref{sec:model} as a common thread, the resulting DMFT/ED approach can be readily generalized to different kinds of jump operators or local Hamiltonians. Special care would be neeeded however in presence of coherent driving fields or for large enough hopping between bosons, favoring the development of a nonequilibrium condensate with a finite value of the local bosonic field. Since we will not focus on these regimes, we will assume the bosons to remain incoherent. For a more general treatment including symmetry-broken phases, we refer the reader to Ref.~\cite{Strand2015,Scarlatella2020}. Finally, since the DMFT approach is formulated for stationary states, we do not discuss any transient effect.

\subsection{DMFT Mapping onto a Markovian Quantum Impurity Model}
\label{sec:dmft_mapping}

In this Section we provide the basic ideas of DMFT for open quantum systems described by a Lindblad master equation of the form given in Sec.~\ref{sec:model}, focusing on the physical interpretation, while referring to Ref.~\cite{Scarlatella2020} for further details.

\begin{figure*}
    \centering
    \includegraphics[scale=0.45]{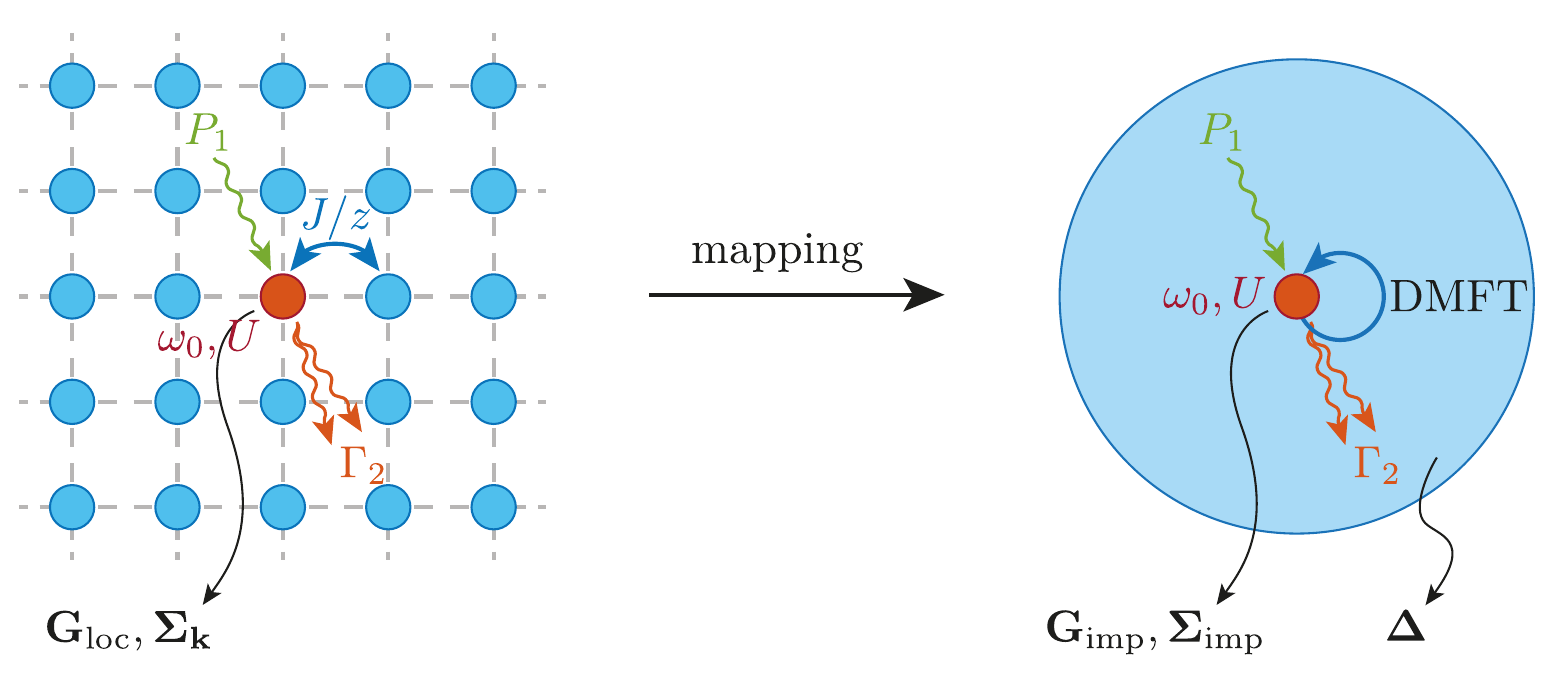}
    \caption{Sketch of the DMFT technique. We start from a translationally-invariant lattice problem, in which each site has nonlinearities --- e.g.\@ a Kerr nonlinearity $U$ and two-particle losses $\Gamma_2$. The lattice sketched here has a nearest-neighbor hopping strength $J/z$, where $z$ is the lattice coordination number. Since the lattice is translationally invariant, we focus on a single site --- marked here in red; at this point, the original lattice problem can be seen as the problem of a single site embedded in some effective bath $\mat{\Delta}$ generated by its interaction with all the other lattice sites. The DMFT technique provides an equation that mathematically connects such bath to the original lattice problem, by discarding spatial correlations in the lattice self-energy $\mat{\Sigma}_{\vec{k}}$. This connection is discussed in Sec.~\ref{sec:dmft_mapping}.}
    \label{fig:dmft_sketch}
\end{figure*}

Dynamical mean-field theory~\cite{Georges1996} realizes a quantum version of the mean-field theory, which becomes exact in the limit of large lattice coordination number or dimensionality. In our implementation, the many-body Lindblad problem in Eqs.~(\ref{eq:bh_model}) and~(\ref{eq:lindblad_term_dissipator}) is mapped onto an auxiliary impurity model characterized by a single site --- called the \emph{impurity} --- with the same local Hamiltonian $\hat{H}_{\mathrm{loc}}(\hat{n})$ and local jump operators $\hat{L}_{\alpha}$ of the original lattice problem\footnote{Since we now have a single site rather than a lattice, we dropped the subscript $\vec{r}$.} (see  Eqs.~(\ref{eq:bh_model}), (\ref{eq:jump_lattice_pump}) and~(\ref{eq:jump_lattice_loss})), embedded into a \emph{linear} non-Markovian quantum bath. The impurity site is representative of an arbitrary site of the original lattice, while the bath describes effectively the interaction of the site with the rest of the lattice.

This idea is mathematically formulated in terms of single-particle Green's functions which, due to the non-equilibrium nature of our problem, are expressed in the Keldysh formalism as $2\times2$ matrices with only two non-trivial components, called the \emph{retarded} and the \emph{Keldysh} components, respectively marked by superscripts $R$ and $K$ in the following~\cite{Kamenev2009}. As a matter of fact, the steady-state lattice Green's function at momentum $\vec{k}$ has the following matrix structure
\begin{equation}
    \mat{G}_{\vec{k}}(\omega) = 
    \left(
    \begin{array}{cc}
        G_{\vec{k}}^K(\omega) & G_{\vec{k}}^R(\omega) \\
        G_{\vec{k}}^A(\omega) & 0 
    \end{array}
    \right)\,,
\end{equation}
where causality implies that the advanced component $G_{\vec{k}}^A$ is related to the retarded component via $G_{\vec{k}}^A = \left( G_{\vec{k}}^R \right)^{\dagger}$, while the Keldysh component is anti-Hermitian as a consequence of its non-equilibrium nature: $G_{\vec{k}}^K = -\left( G_{\vec{k}}^K \right)^{\dagger}$.

The key quantity is the impurity Green's function $\mat{G}_{\mathrm{imp}}$:
\begin{equation}
    \mat{G}_{\mathrm{imp}}^{-1}(\omega) = \mat{g}_{0,\mathrm{imp}}^{-1}(\omega) - \mat{\Delta}(\omega) - \mat{\Sigma}_{\mathrm{imp}}(\omega).
    \label{eq:G_imp_general}
\end{equation}
In the expression above, $\mat{g}_{0,\mathrm{imp}}$ is the trivial Green's function of a non-interacting ($U=\Gamma_2=0$, marked by the subscript $0$) impurity site when disconnected (marked by the lowercase $g$) from the bath. Its explicit expression, while not needed here, can be read off from Eqs.~(\ref{eq:DeltaR}) and~(\ref{eq:DeltaK}) in the limit of a single bath site and in absence of any single-particle loss. The coupling between the impurity site and the auxiliary bath is instead encoded into $\mat{\Delta}$, called the \emph{bath hybridization function}, which is known analytically, while the self-energy $\mat{\Sigma}_{\mathrm{imp}}$ encodes all the nonlinearities resulting both from local processes ($U,\,\Gamma_2$) as well as from the coupling between the impurity and the auxiliary bath.

The DMFT mapping is enforced by identifying the impurity Green's function in Eq.~(\ref{eq:G_imp_general}) with the \emph{local} Green's function of the lattice many-body problem. Namely, if we write the local part of the lattice Green's function
\begin{equation}
    \mat{G}_{\mathrm{loc}}(\omega) = \sum_{\vec{k}} \left( \mat{G}_{0,\vec{k}}^{-1}(\omega) - \mat{\Sigma}_{\vec{k}}(\omega) \right)^{-1} \,,
    \label{eq:lattice_green_local}
\end{equation}
where $\mat{G}_{0,\vec{k}}$ is the $\vec{k}$-dependent non-interacting ($U=\Gamma_2=0$) Green's function of the lattice and $\mat{\Sigma}_{\vec{k}}$ is its $\vec{k}$-dependent lattice self-energy, within DMFT we have
\begin{equation}
    \mat{G}_{\mathrm{loc}}(\omega) = \mat{G}_{\mathrm{imp}}(\omega)
    \qquad\text{and}\qquad
    \mat{\Sigma}_{\vec{k}}(\omega) = \mat{\Sigma}_{\mathrm{imp}}(\omega)\,,
    \label{eq:dmft_equalities}
\end{equation}
i.e.\@ the lattice self-energy becomes $\vec{k}$-independent and equal to the impurity self-energy, and the impurity Green's function matches the local Green's function of the lattice --- see the sketch in Fig.~\ref{fig:dmft_sketch}. 

The core of an actual DMFT calculation is the evaluation of $\mat{G}_{\mathrm{imp}}(\omega)$ or equivalently $\mat{\Sigma}_{\mathrm{imp}}(\omega)$, which requires some numerical or analytical method which is usually called an \emph{impurity solver}. Enforcing the first condition of (\ref{eq:dmft_equalities}) implies that the solution must be self-consistent, since the hybridization function has to be adjusted in order to satisfy the DMFT condition. As a matter of fact, this is realized by an iterative self-consistent loop that we outline in Sec. \ref{sec:dmft} together with our implementation. Note that, in a non-equilibrium setting, the retarded and Keldysh components are not necessarily related by the fluctuation-dissipation theorem; therefore, the consistency of the DMFT mapping is ensured by imposing Eq.~(\ref{eq:dmft_equalities}) for both the retarded and the Keldysh component.

The self-consistent solution of the quantum impurity model is the main technical challenge behind any DMFT approach. This is particularly true in the case of open quantum many-body systems, where the resulting impurity model in Eq.~(\ref{eq:G_imp_general}) contains interactions, non-equilibrium effects due to the local jump operators and the coupling to a large frequency-dependent bath. Below, we present our implementation of an impurity solver for DMFT based on exact diagonalization of the associated Lindblad superoperator.

\subsection{Lindblad Representation of the DMFT Bath and Exact Diagonalization}
\label{sec:lindblad_dmft_ed}

In order to solve the effective impurity model using an exact-diagonalization solver, we have to represent the non-Markovian bath as a finite (and numerically affordable) matrix. For equilibrium DMFT problems, involving either fermions or bosons, this can be readily done by introducing a set of non-interacting ($U=0$) auxiliary sites whose energies and couplings with the impurity site are chosen in order to reproduce the spectral properties of the bath as well as possible~\cite{Georges1996}.

For generic out-of-equilibrium problems, however, one needs to take into account not only the spectrum of the bath but also its occupation, since the FDT does not hold \textit{a priori}. The occupation of the auxiliary sites can be controlled by providing them with dissipative processes that exchange single particles between some Markovian reservoirs and the auxiliary sites themselves, at rates to be determined so to reproduce the occupation properties of the bath; these are precisely the kind of dissipative processes described by a Lindblad equation with single-particle jump operators. This argument was used in Refs.~\cite{Arrigoni2013,Dorda2014}, which studied steady-state transport in a correlated electronic system with DMFT, to suggest a Lindblad representation of the non-equilibrium DMFT bath. We can therefore follow this approach, with the additional difference that here we also have driving and dissipation on the impurity site and that we consider a bosonic version of the problem.

We start by writing down an effective bosonic Anderson impurity model (AIM) whose dynamics is described by a Lindblad master equation
\begin{equation}
    \frac{d}{dt}\hat{\rho} = \doublehat{\mathcal{L}}\hat{\rho} = \doublehat{\mathcal{L}}_H\hat{\rho} + \doublehat{\mathcal{L}}_D\hat{\rho}\,,
    \label{eq:AIM_Hermitian_evolution}
\end{equation}
where
\begin{equation}
    \doublehat{\mathcal{L}}_H\hat{\rho} = -i\left[\hat{H},\hat{\rho}\right]
    \label{eq:AIM_dissipative_evolution}
\end{equation}
is the coherent part of the evolution with a Hamiltonian 
\begin{align}
    \hat{H} 
    &= \omega_0\hat{a}^{\dagger}\hat{a} + U\hat{a}^{\dagger}\hat{a}\hat{a}^{\dagger}\hat{a} \nonumber \\
    &\phantom{={}} 
    + \sum_{n=1}^{N_B} \Big\lbrace \omega_n\hat{b}^{\dagger}_n\hat{b}_n
    + \nu_n\hat{b}^{\dagger}_n\hat{a} + \nu_n^*\hat{a}^{\dagger}\hat{b}_n \Big\rbrace
    \label{eq:AIM_Hamiltonian}
\end{align}
describing an interacting bosonic impurity, with creation (annihilation) operator $\hat{a}^{\dagger}$ ($\hat{a}$), linearly coupled to a free bosonic bath with energies $\omega_{n}$, creation (annihilation) operators $\hat{b}^{\dagger}_n$ ($\hat{b}_n$) and coupling strengths to the impurity $\nu_n$, which in the following we take to be real ($\nu_n = \nu_n^*$).

\begin{figure}
    \centering
    \includegraphics[scale=0.45]{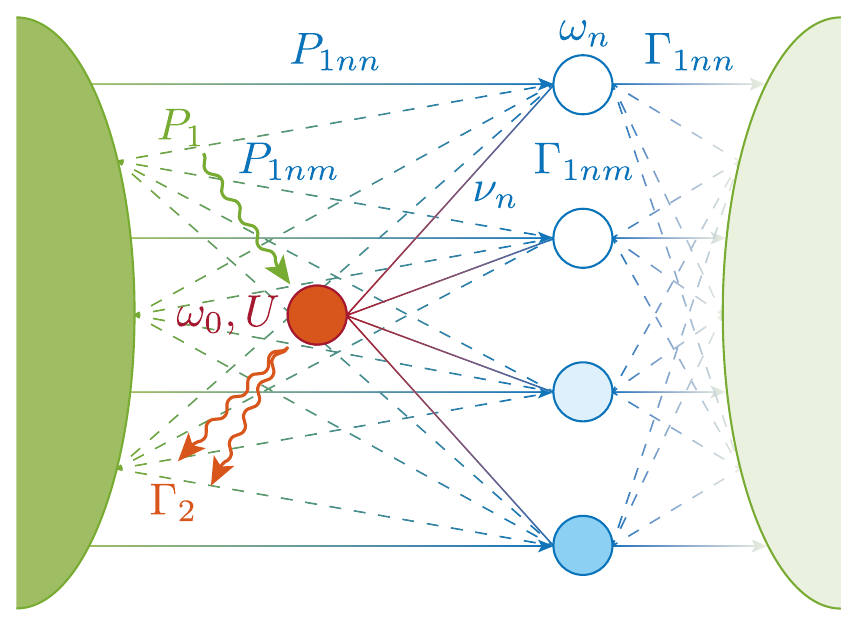}
    \caption{Sketch of a non-equilibrium AIM used to model a single driven-dissipative lattice site plus effective bath in DMFT. The impurity (central red site) is the only site having nonlinearities, and it is connected to the auxiliary bath sites via some hopping amplitudes $\nu_n$, where $n=1,\ldots,N_N$ indexes the bath sites. The bath sites have energies $\omega_n$ and are connected to a completely full Markovian environment (dark green blob on the left), which provides single-particle pumping, and to a completely empty Markovian environment (light green blob on the right), which provides single-particle losses. The connection is visualized in terms of links corresponding to the Lindblad coefficients, with $P_{1mn}$ and $\Gamma_{1mn}$ representing the coefficients matrices of the single-particle pump and loss processes, respectively. The diagonal elements of the coefficient matrices correspond to solid lines, while the off-diagonal elements to dashed lines. An analogous AIM was used in Refs.~\cite{Arrigoni2013,Dorda2014} to map a Hermitian lattice model in the context of non-equilibrium fermionic DMFT.}
    \label{fig:aim_amea_rh}
\end{figure}

For what concerns the Lindblad dissipator $\doublehat{\mathcal{L}}_D$, we note that there is a certain freedom in the choice of the jump operators of the bath. In fact, since the auxiliary cavities are a mere fitting tool, any linear combination of $\{\hat{b}^{\dagger}_n\}$ (or $\{\hat{b}_n\}$) which is consistent with the properties of the Lindblad equation is still a valid jump operator. Equivalently, modulo a unitary transformation, we can instead allow for off-diagonal Lindblad couplings among fixed jump operators $\{\hat{b}^{\dagger}_n\}$ (or $\{\hat{b}_n\}$), so that the Lindblad dissipator can be generically written as --- see also the sketch in Fig.~\ref{fig:aim_amea_rh}:
\begin{align}
    \doublehat{\mathcal{L}}_D\hat{\rho} 
    &= 2\Gamma_2 \left( \hat{a}\hat{a} \hat{\rho} \hat{a}^{\dagger}\hat{a}^{\dagger}
        - \frac{1}{2}\left\lbrace \hat{a}^{\dagger}\hat{a}^{\dagger}\hat{a}\hat{a},\hat{\rho} \right\rbrace \right) \nonumber \\
    &\phantom{={}} 
    + 2P_1 \left( \hat{a}^{\dagger} \hat{\rho} \hat{a}
        - \frac{1}{2}\left\lbrace \hat{a}\hat{a}^{\dagger},\hat{\rho} \right\rbrace \right) \nonumber \\
    &\phantom{={}} 
    + 2 \sum_{n,m=1}^{N_B}
        \Bigg\lbrace 
        \Gamma_{1mn} \left( \hat{b}_n \hat{\rho} \hat{b}_m^{\dagger}
            - \frac{1}{2}\left\lbrace \hat{b}_m^{\dagger}\hat{b}_n,\hat{\rho} \right\rbrace \right) \nonumber \\
    &\phantom{{}=} \phantom{+ 2 \sum_{n,m=0}^{N_B} \Bigg\lbrace} \nhphantom{+{}} 
        + P_{1mn} \left( \hat{b}_m^{\dagger} \hat{\rho} \hat{b}_n
            - \frac{1}{2}\left\lbrace \hat{b}_n\hat{b}_m^{\dagger},\hat{\rho} \right\rbrace \right)
        \Bigg\rbrace\,.
    \label{eq:AIM_Lindblad_dissipator}
\end{align}
In this expression we see both dissipative processes for the impurity, inherited from the local jump operators (\ref{eq:jump_lattice_pump})--(\ref{eq:jump_lattice_loss}) of the original lattice model, as well as single-particle drive and losses for the auxiliary bath sites, parametrized by the coefficients $\Gamma_{1mn}$ and $P_{1mn}$, with $n,m=1,\ldots,N_B$.

We emphasize that the impurity parameters in the model, namely the local frequency $\omega_0$, the interaction $U$, and the rate of single-particle drive $P_1$ and two-particle losses $\Gamma_2$, are identical to those used in Eq.~(\ref{eq:lindblad_term_dissipator}) for a generic lattice site, by construction of the DMFT mapping. On the other hand, the parameters that characterize the auxiliary sites and their coupling to the impurity, as well as their dissipation, are chosen in such a way to represent the DMFT bath.

In particular, for the diagonal Lindblad couplings we employ in the numerical calculations --- i.e.\@ for $\Gamma_{1nm} = \Gamma_{1nn}\delta_{nm}$ and $P_{1nm} = P_{1nn}\delta_{nm}$, with $n,m = 1,\ldots,N_B$, and upon integrating the degrees of freedom of the bath, we obtain the following expressions for the retarded/Keldysh components of the bath hybridization function $\mat{\Delta}$ appearing in Eq.~(\ref{eq:G_imp_general})~\cite{Secli2021b}:
\begin{equation}
    \Delta^R(\omega)
    = \sum_{n=1}^{N_B} \nu_n^2 
    \frac{1}{\omega-\omega_n + i(\Gamma_{1nn}-P_{1nn})}
    \label{eq:DeltaR}
\end{equation}
and 
\begin{equation}
    \Delta^K(\omega)
    = -2i \sum_{n=1}^{N_B} \nu_n^2 
    \frac{\Gamma_{1nn}+P_{1nn}}{(\omega-\omega_n)^2 + (\Gamma_{1nn}-P_{1nn})^2} \,,
    \label{eq:DeltaK}
\end{equation}
representing respectively the spectral function and the occupation function of the bath. Note, in particular, that their general structure is that of a weighted sum of Lorentzians; each auxiliary site contributes a single Lorentzian in each component of $\mat{\Delta}$, centered at the site bare frequency $\omega_n$ and with a HWHM equal to the bare effective loss rate $\Gamma_{1nn}-P_{1nn}$ at that site.

Having a discretized representation of the DMFT bath, we can finally solve the AIM described by Eq.~(\ref{eq:AIM_Hermitian_evolution}) in order to numerically obtain the impurity Green's function $\mat{G}_{\mathrm{imp}}$, whose retarded/Keldysh components are respectively
\begin{align}
    G_{\mathrm{imp}}^R(\omega) &= -i \int dt\, e^{i\omega t} \theta(t)\Braket{\left[\hat{a}(t), \hat{a}^{\dagger}(0)\right]}
    \label{eq:G_imp_R_defs} \,,\\
    G_{\mathrm{imp}}^K(\omega) &= -i \int dt\, e^{i\omega t} \Braket{\left\lbrace \hat{a}(t), \hat{a}^{\dagger}(0)\right\rbrace} \,.
    \label{eq:G_imp_K_defs}
\end{align}
As illustrated in detail in Appendix\nobreakspace \ref{app:aimdiag}, our impurity solver is based on an exact-diagonalization scheme. The starting point is the vectorization of the Lindblad equation, in which the density matrix $\hat{\rho}$ (the Lindbladian superoperator $\doublehat{\mathcal{L}}$) is represented as a vector $\ket{\rho}$ (a matrix $\hat{\mathcal{L}}$) in Fock space:
\begin{equation}
    \frac{d}{dt}\hat{\rho} = \doublehat{\mathcal{L}}\hat{\rho}
    \qquad\Longrightarrow\qquad
    \frac{d}{dt}\ket{\rho} = \hat{\mathcal{L}}\ket{\rho} \,.
\end{equation}
In this representation, $\hat{\mathcal{L}}$ is a general (i.e.\@ non-symmetric) complex-valued matrix; therefore, it is necessary to perform a two-sided diagonalization that yields both left and right eigenvectors, respectively called $\bra{l_{\alpha}}$ and $\ket{r_{\alpha}}$, as well as complex eigenvalues $\mathcal{L}_{\alpha}$. The retarded/Keldysh components of the impurity Green's function, Eqs.~(\ref{eq:G_imp_R_defs}) and~(\ref{eq:G_imp_K_defs}), can then be expressed as a function of the eigenvectors and eigenvalues resulting from the diagonalization of $\hat{\mathcal{L}}$ --- see Eqs.~(\ref{eq:G_R_KL_W}) and~(\ref{eq:G_K_KL_W}) and Appendix\nobreakspace \ref{app:aimdiag} for further details.

\subsection{The DMFT/ED loop}
\label{sec:dmft_loop}

Having described the DMFT mapping in Sec.~\ref{sec:dmft_mapping} and our specific discrete representation of the bath in Sec.~\ref{sec:lindblad_dmft_ed}, we can now conclude our overview of the DMFT approach by discussing the full DMFT loop, including the self-consistency condition.

As already noted in Sec.~\ref{sec:dmft_mapping}, the DMFT mapping is obtained by enforcing the equalities (\ref{eq:dmft_equalities}). These, in practice, with our discretized representation of the DMFT bath, are self-consistently satisfied by performing the following \emph{DMFT loop}:
\begin{enumerate}
    \item\label{item:dmft_bath_01} \underline{Initial guess}.\\
    Start from a guess for the bath parameters $\omega_n$, $\nu_n$, $\Gamma_{1nn}$, and $P_{1nn}$, with $n=1,\ldots,N_B$, which corresponds to a guess for the bath hybridization functions $\Delta^{R/K}(\omega)$ given by Eqs.~(\ref{eq:DeltaR}) and~(\ref{eq:DeltaK}).
    \item\label{item:dmft_bath_02} \underline{From $\mat{\Delta}$ to $\mat{G}_{\mathrm{imp}}$}.\\
    Given the bath hybridization functions $\Delta^{R/K}$, corresponding to a set of bath parameters for the AIM, obtain the impurity Green's functions $G^{R/K}_{\mathrm{imp}}$ by diagonalizing the AIM via ED --- see Appendix\nobreakspace \ref{app:aimdiag} and Eqs.~(\ref{eq:G_R_KL_W}) and~(\ref{eq:G_K_KL_W}).
    \item\label{item:dmft_bath_03} \underline{From $\mat{G}_{\mathrm{imp}}$ to $\mat{\Delta}^{\mathrm{(new)}}$}.\\
    Given $G^{R/K}_{\mathrm{imp}}$, calculate a new bath hybridization function $\mat{\Delta}^{\mathrm{(new)}}$. This is the step that involves the original lattice, i.e.\@ the step in which we actually enforce the DMFT equalities (\ref{eq:dmft_equalities}).
    \begin{enumerate}
        \item\label{item:dmft_bath_03-01} Obtain the impurity self-energy  as: $\mat{\Sigma}_{\mathrm{imp}} = \mat{g}_{0,\mathrm{imp}}^{-1} - \mat{\Delta} - \mat{G}_{\mathrm{imp}}^{-1}$.
        \item\label{item:dmft_bath_03-02} Perform the DMFT approximation on the self-energy: $\mat{\Sigma}_{\vec{k}} = \mat{\Sigma}_{\mathrm{imp}}$.
        \item\label{item:dmft_bath_03-03} Compute the local lattice Green's function $\mat{G}_{\mathrm{loc}}$ via Eq.~(\ref{eq:lattice_green_local}).
        \item\label{item:dmft_bath_03-04} Impose the DMFT condition to get a new impurity Green's function: $\mat{G}_{\mathrm{imp}}^{\mathrm{(new)}} = \mat{G}_{\mathrm{loc}}$.
        \item\label{item:dmft_bath_03-05} Numerically compute the new bath hybridization function as: $\mat{\Delta}^{\mathrm{(new)}} = \mat{g}_{0,\mathrm{imp}}^{-1} - \left(\mat{G}_{\mathrm{imp}}^{\mathrm{(new)}}\right)^{-1} - \mat{\Sigma}_{\mathrm{imp}}$.
    \end{enumerate}
    \item\label{item:dmft_bath_02-01} \underline{Fit of $\mat{\Delta}^{\mathrm{(new)}}$}.\\
    Find new bath parameters such that the bath hybridization functions computed via Eqs.~(\ref{eq:DeltaR}) and~(\ref{eq:DeltaK}) are as close as possible to the given numerical $\mat{\Delta}^{\mathrm{(new)}}$.
    This is done, as we will see, via a fitting procedure that minimizes a suitable distance between the hybridization functions.
    \item\label{item:dmft_bath_04} \underline{Convergence test}.\\
    If the distance between the new bath hybridization functions and the former ones is less than the specified tolerance, i.e.\@ if $\left\vert\left\vert \mat{\Delta}^{\mathrm{(new)}} - \mat{\Delta} \right\vert\right\vert < \delta_{\mat{\Delta}}$, stop. Otherwise, set $\mat{\Delta} = \mat{\Delta}^{\mathrm{(new)}}$, go back to step \ref{item:dmft_bath_02} and iterate until convergence.
\end{enumerate}

In some cases, the step \ref{item:dmft_bath_03} involving lattice quantities can be greatly simplified. On a Bethe lattice\index{Bathe lattice}, in particular, the self-consistency relation simply reads \cite{Georges1996,Strand2015}:
\begin{equation}
    \mat{\Delta} =  \frac{J^2}{z} \cdot \mat{G}_{\mathrm{imp}} \,.
    \label{eq:Bethe_SCF}
\end{equation}

The last point to address is how we perform the fitting step in \ref{item:dmft_bath_02-01}. First, in analogy with the treatment in \cite{Dorda2014}, we define a distance $\chi(\mat{\Delta}_1,\mat{\Delta}_2)$ between two generic hybridization functions $\mat{\Delta}_1$, $\mat{\Delta}_2$ as:
\begin{equation}
    \chi(\mat{\Delta}_1,\mat{\Delta}_2) = \sum_{\alpha=R,K}
    \int_{-\infty}^{+\infty} d\omega\, W^{\alpha}(\omega)
    \left\lvert \Delta_1^{\alpha}(\omega) - \Delta_2^{\alpha}(\omega) \right\rvert^n \,,
\end{equation}
where $W^{\alpha}(\omega)$ is a positive-valued weight function which can be typically set to a constant $W^{\alpha}(\omega) \equiv W$, $\vert \cdot \vert$ is the complex norm, and $n$ can be typically fixed to $2$.

The fitting procedure is then achieved by performing a multidimensional numerical minimization of the function $\chi(\mat{\Delta}_{\mathrm{target}},\mat{\Delta})$, where $\mat{\Delta}_{\mathrm{target}}$ is the \emph{numerical} data we want to fit, while $\mat{\Delta}$ are the \emph{analytical} fitting functions in Eqs.~(\ref{eq:DeltaR}) and~(\ref{eq:DeltaK}). By default, our numerical code \cite{Secli2022} uses the L-BFGS-B minimizer provided by \texttt{Cpp\-Optimization\-Library} \cite{Wieschollek2016}. The L-BFGS-B algorithm is in fact able to handle simple box constraints, which in our case are used to ensure that the coefficient matrices in the Lindblad equation for the AIM are positive semi-definite.

\section{Steady-State Quantum Zeno Regime of the Bose--Hubbard Model}
\label{sec:zeno}

In this Section, we present our DMFT results on the driven-dissipative BH model introduced in Sec.~\ref{sec:model}, focusing in particular on the regime of strong two-particle losses. To this extent, we briefly recall the results known in the literature for the case of a purely dissipative (no pump) problem, studied in Refs.~\cite{Garcia-Ripoll2009,Rossini2021}, and then we introduce a weak single-particle pump to stabilize a finite-density stationary state. We discuss the signatures of the QZE in the local properties of the model, in particular in the occupation and in the local spectral and correlation functions, and we show how our ED impurity solver qualitatively reproduces the results obtained in Ref.~\cite{Scarlatella2020} with NCA. Finally, we present further insights on the origin of the QZE within DMFT by looking at the structure of the DMFT bath in this regime, and we provide evidence for an effective dimer model to capture this physics.

\subsection{Decay to the Vacuum and Emergence of Hard-Core Bosons}
\label{sec:hard_core_bosons}

We begin our discussion by considering the dissipative Bose--Hubbard model obtained from Eqs.~(\ref{eq:bh_model}) and~(\ref{eq:lindblad_term_dissipator}) by removing the incoherent pump, which in the long-time limit ultimately corresponds to the complete depletion of the system. In the regime of large dissipation, $\Gamma_2 \gg J$, depletion is slow and the dynamics of the system is described by an effective hard-core boson problem, following Ref.~\cite{Garcia-Ripoll2009}. In fact, when $\Gamma_2$ is the dominant energy scale, all states with two or more bosons per site acquire a finite (and large) decay and do not contribute to the long-time limit of the system. As was shown in Ref.~\cite{Garcia-Ripoll2009}, the manifold controlling the long-time limit of the model spans therefore states containing empty and singly occupied sites only, i.e. the system can be mapped into an effective hard-core boson problem with effective Hamiltonian
\begin{equation}
    \hat{H}^{\mathrm{eff}}
    = - \frac{J}{z} \sum_{\braket{\vec{r},\,\vec{r}'}} \hat{c}_{\vec{r}}^{\dagger}\hat{c}_{\vec{r}'} - J_2^{\mathrm{eff}} \sum_{\vec{r}} \hat{C}_{\vec{r}}^{\dagger}\hat{C}_{\vec{r}},
    \label{eq:bh_model_eff_Hamiltonian}
\end{equation}
where $\hat{c}_{\vec{r}}$, $\hat{c}_{\vec{r}}^{\dagger}$ are \emph{hard-core} bosonic operators\index{hard-core bosons}, i.e.\@ with a constraint enforcing a maximum occupation of one boson per site:
\begin{equation}
    \hat{c}_{\vec{r}} = \ket{0}_{\vec{r}}\bra{1}_{\vec{r}},
    \qquad
    \hat{c}_{\vec{r}}^{\dagger} = \ket{1}_{\vec{r}}\bra{0}_{\vec{r}}.
    \label{eq:bh_model_eff_c}
\end{equation}
The second term in Eq.~(\ref{eq:bh_model_eff_Hamiltonian}) is instead written in terms of two-particle operators
\begin{equation}
    \hat{C}_{\vec{r}} = \hat{c}_{\vec{r}}\sum_{\vec{r}' : \braket{\vec{r},\,\vec{r}'}}\hat{c}_{\vec{r}'}\,,
    \label{eq:bh_model_eff_C}
\end{equation}
where the sum over $\vec{r}'$ is carried over the first nearest-neighbors of $\vec{r}$, that destroy \emph{pairs} of bosons in neighboring sites. The quantity
\begin{equation}
    J_2^{\mathrm{eff}} = \left(\frac{J}{z}\right)^2 \frac{U}{U^2+\Gamma_2^2}
    \label{eq:bh_model_eff_J2}
\end{equation}
is the effective energy associated to these pairs.  

In addition to these hopping processes, the hard-core bosons are still subject to a small residual dissipation that ultimately empties out the lattice. This is a virtual dissipative process that is obtained whenever a photon in a singly occupied bosonic site hops to another neighboring singly occupied site, and forms a doublon which is then quickly ejected from the system due to the fast two-particle losses. In other words, the local two-particle dissipator in the Lindblad equation is replaced by an effective non-local two-particle dissipator
\begin{align}
    \doublehat{\mathcal{L}}_D^{\mathrm{eff}}\hat{\rho}^{\mathrm{BH}} = 2 \sum_{\vec{r}}
        &\Gamma_2^{\mathrm{eff}} \left( \hat{C}_{\vec{r}} \hat{\rho}^{\mathrm{BH}} \hat{C}_{\vec{r}}^{\dagger}
        - \frac{1}{2}\left\lbrace \hat{C}_{\vec{r}}^{\dagger}\hat{C}_{\vec{r}},\hat{\rho}^{\mathrm{BH}} \right\rbrace \right)\,,
    \label{eq:bh_model_eff_dissipator}
\end{align}
where
\begin{equation}
    \Gamma_2^{\mathrm{eff}} = \left(\frac{J}{z}\right)^2 \frac{\Gamma_2}{U^2+\Gamma_2^2}
    \label{eq:bh_model_eff_G2}
\end{equation}
is the effective decay rate of pairs of bosons on neighboring sites. The peculiarity of the effective dissipation is that it is mediated by the original tunneling amplitude $J$: as $J$ increases, more and more bosons on neighboring sites are discarded from the lattice. Even more interestingly, for $\Gamma_2 < U$ the effective dissipation increases with the bare dissipation $\Gamma_2$, but only up to a maximum at $\Gamma_2 = U$; for $\Gamma_2 > U$, instead, the effective dissipation \emph{decreases} as the bare dissipation $\Gamma_2$ is increased. Eventually, at $\Gamma_2 \gg U$, the effective dissipation becomes
\begin{equation}
    \Gamma_2^{\mathrm{eff}} \underset{\Gamma_2 \gg U}{\approx} \left(\frac{J}{z}\right)^2 \frac{1}{\Gamma_2}\,,
\end{equation}
thus it decreases as $\Gamma_2^{-1}$. This regime where the effective dissipation shows a seemingly paradoxical behavior, experimentally observed with ultracold gases \cite{Syassen2008,Yan2013,Tomita2017}, has been called the \emph{quantum Zeno regime}\index{quantum Zeno regime} for its analogy with Zeno’s arrow paradox \cite{Hardie1984}.
We emphasize that the original works on the QZE focus on the effect of frequent projective measurements on the unitary dynamics of an isolated system and the consequent Zeno localization~\cite{Misra1977,Beige2000,Peres1980,Itano1990}. In the context of open quantum systems described by a Lindblad master equation, the analogy is best understood by unravelling the master equation into quantum trajectories and interpreting the resulting quantum jumps as the effect of a measurement click~\cite{Wiseman2009}. In this context, the QZE corresponds to the regime of strong dissipation, when the system is dynamically projected in the dark sector of the Lindblad dissipator~\cite{Garcia-Ripoll2009,Rossini2021} rather than exploring the full Hilbert space. In the present problem this sector, composed of singly occupied and empty sites, can still feature a non-trivial unitary dynamics and a small residual hopping-induced dissipation~(\ref{eq:bh_model_eff_G2}). The hallmark of the QZE is therefore the scaling of the effective dissipation as $1/\Gamma_2$.

Because of their origin in terms of virtual processes, the hopping-induced dissipative processes described by Eq.~(\ref{eq:bh_model_eff_dissipator}) are completely neglected by simple approaches such as the Gutzwiller mean-field theory and, even in the context of DMFT, they cannot be handled using approximate solvers such as the Hubbard-I approximation~\cite{Scarlatella2020}. On the other hand, our non-perturbative implementation of DMFT is not limited in this regard, and it is an ideal tool to explore this regime. We shall also see that our ED implementation provides an intriguing and insightful mapping onto an effective two-site model.

\subsection{Steady-State Quantum Zeno Regime}
\label{sec:steadystatezeno}

In Sec.~\ref{sec:hard_core_bosons}, we considered the BH model in absence of any pump. Here, instead, we take the same approach of Ref.~\cite{Scarlatella2020} by including a small single-particle pumping $P_1 \ll \Gamma_2$ in the lattice in order not to have an empty stationary state. This is a key difference compared to the effective model in Eqs.~(\ref{eq:bh_model_eff_Hamiltonian}) to~(\ref{eq:bh_model_eff_G2}), derived in \cite{Garcia-Ripoll2009} for a lattice without a pumping mechanism.  

\begin{figure}
    \centering
    \includegraphics[scale=0.45]{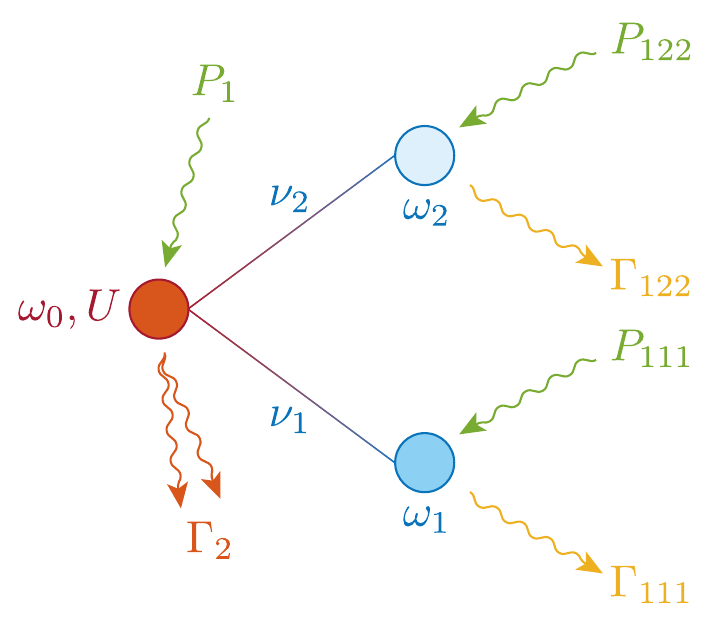}
    \caption{Effective AIM used for the numerical simulations in Sec.~\ref{sec:zeno}. The impurity site, representing a generic site of the original lattice, is depicted on the left and has a bare frequency $\omega_0$ and a Kerr nonlinearity $U$; the bath sites, depicted on the right, do not have Kerr nonlinearities and have bare frequencies $\omega_1$, $\omega_2$ and couplings with the impurity $\nu_1$, $\nu_2$ that are treated as fitting parameters. Every site has single-particle pumping, with $P_1$ fixed and $P_{111}$, $P_{122}$ fitting parameters. The impurity site has two-particle losses $\Gamma_2$, while the auxiliary bath sites have single-particle losses $\Gamma_{111}$ and $\Gamma_{122}$, again treated as fitting parameters.}
    \label{fig:aim_zeno}
\end{figure}

We start performing our DMFT simulations on a Bethe lattice\footnote{A variety of DMFT Hamiltonian studies have demonstrated a strong similarity between the results on the Bethe lattice and on other lattices. Therefore, we can expect a similar generality also for our results, while retaining the numerical benefits of using the simplified self-consistency relation (\ref{eq:Bethe_SCF}).} with coordination number $z=6$. We fix $\omega_0=1$ and we set the Kerr-like nonlinearity to $U=10$, which is large when compared to $\omega_0$, so to have enough separation --- equal to $2U$ for a single disconnected cavity --- between the peaks in the spectral function. In the following, we measure all the energy scales in units of $U$. As for the hopping strength $J$, since $\Gamma_2^{\mathrm{eff}}$ is proportional to $J^2$, decreasing/increasing $J$ is expected to result in a less/more prominent Zeno effect. We then explore values of $\Gamma_2/U$ from $0.25$ to $1.50$ while we fix $P_1/U=0.01$ in the following. We consider a minimal impurity model with two bath sites connected to the impurity site itself --- see Fig.~\ref{fig:aim_zeno}; as we are going to discuss below, this turns out to be a sufficient representation of the system in the Zeno regime. Since we are focusing on the Zeno regime, where multiple bosonic occupation is suppressed, we can choose quite a small Hilbert space cutoff $N_{\mathrm{cutoff},0}$ on the impurity site in the numerical calculations; in the following, we fixed $N_{\mathrm{cutoff},0} = 5$. As for the Hilbert space cutoff on the bath sites, instead, we have to take some extra care; these sites reproduce the presence of the original lattice itself, but while the single lattice site may mainly occupy the states $\ket{0}$ and $\ket{1}$, this is not necessarily true for a site that mimics the presence of multiple lattice sites. In the following, we fixed $N_{\mathrm{cutoff},1} = N_{\mathrm{cutoff},2} = 7$; we discuss the suitability of this choice below.

\begin{figure}[tbp]
    \centering
    \mbox{\hspace*{-12pt}\includegraphics[scale=0.5]{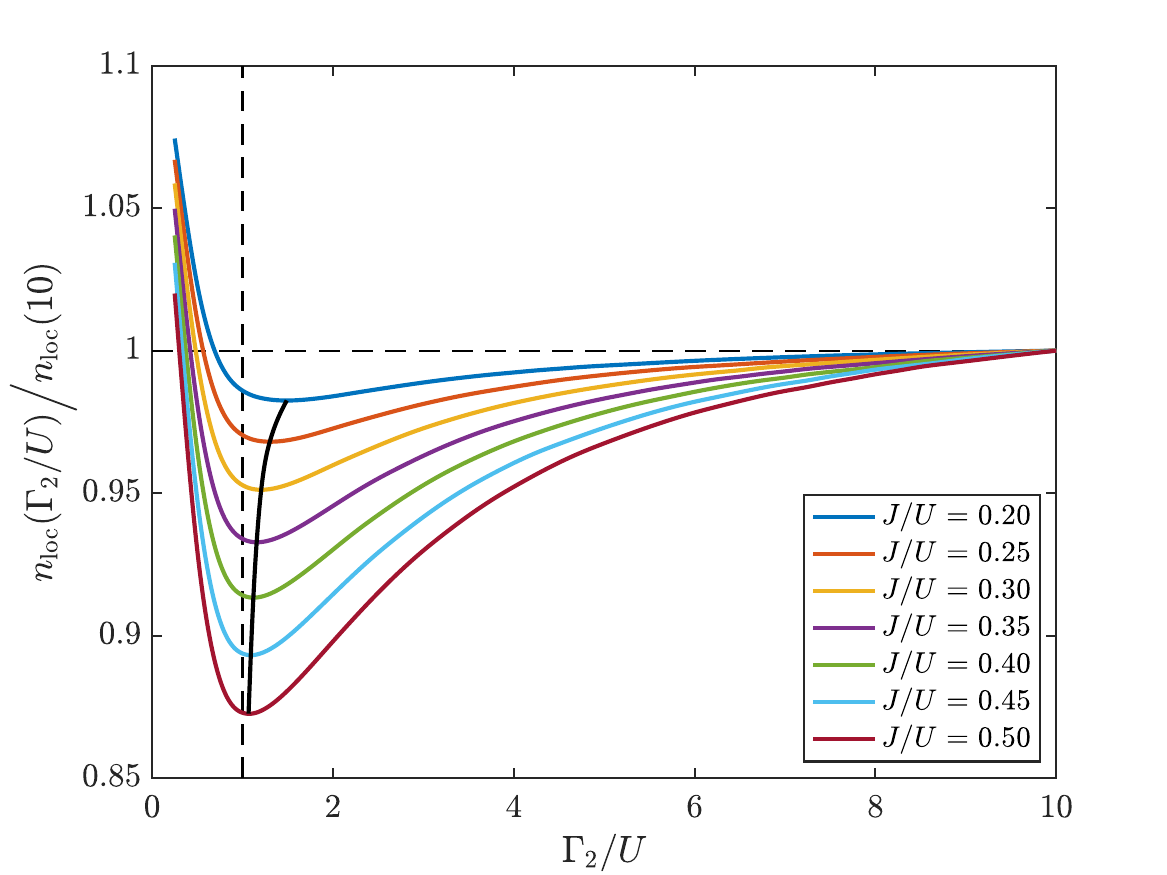}}
    \caption{Mean occupation $n_{\mathrm{loc}} = \braket{\hat{a}^{\dagger}\hat{a}}$ at DMFT self-consistency as a function of $\Gamma_2/U$, normalized to its value at $\Gamma_2/U = 10$; different colors correspond to different values of $J/U$. The horizontal and vertical dashed black lines mark the values $n_{\mathrm{loc}}(\Gamma_2/U)/n_{\mathrm{loc}}(10) = 1$ and $\Gamma_2/U = 1$, respectively; the solid black line marks the occupation minimum for different values of $J/U$. Parameters: as in Fig.~\ref{fig:zeno_emergence_pSS}, except for $J/U=[0.2,0.5]$ and $\Gamma_2/U = [0.25,10.0]$.}
    \label{fig:zeno_meanoccupations}
\end{figure}

We start discussing the signatures of the steady-state quantum Zeno (QZ) regime in the local properties of the BH model, in particular the on-site occupation $n_{\mathrm{loc}}$. This can can be calculated from the steady-state density matrix $\hat{\rho}_{\mathrm{ss}}$ as $n_{\mathrm{loc}} = \braket{\hat{a}^{\dagger}\hat{a}} = \Tr\left(\hat{a}^{\dagger}\hat{a}\hat{\rho}_{\mathrm{ss}}\right)$, and it is shown in Fig.~\ref{fig:zeno_meanoccupations} over a broad range of $\Gamma_2/U$ and $J/U$, normalized to the latest available value in the deep Zeno regime; we show only a selection of $J/U$ values due to data visualization constraints.

The crossover between a regime in which the steady-state occupation decreases and a regime in which it increases does not occur exactly at $\Gamma_2/U = 1$ for any value of $J/U$, as it happens instead for the effective two-particle losses (\ref{eq:bh_model_eff_G2}). The reason, aside from the introduction of a small incoherent pump not present in the effective model of Eqs.~(\ref{eq:bh_model_eff_Hamiltonian}) to~(\ref{eq:bh_model_eff_G2}) and which contributes to the steady-state occupation, is that we also have a first-order contribution in Eq.~(\ref{eq:bh_model_eff_Hamiltonian}), which scales with $J/z$, while the second order contribution responsible for the Zeno effect scales as $(J/U)^2$. The first order contribution is responsible for the shift of the minimum of the steady-state occupation; however, when $J$ is increased, the second order contribution becomes progressively dominant, resulting in a convergence of the stationary point of the steady-state occupation towards $\Gamma_2/U = 1$ --- see the solid black line in Fig.~\ref{fig:zeno_meanoccupations}.

\begin{figure}[tbp]
    \centering
    \mbox{\hspace*{-12pt}\includegraphics[scale=0.5]{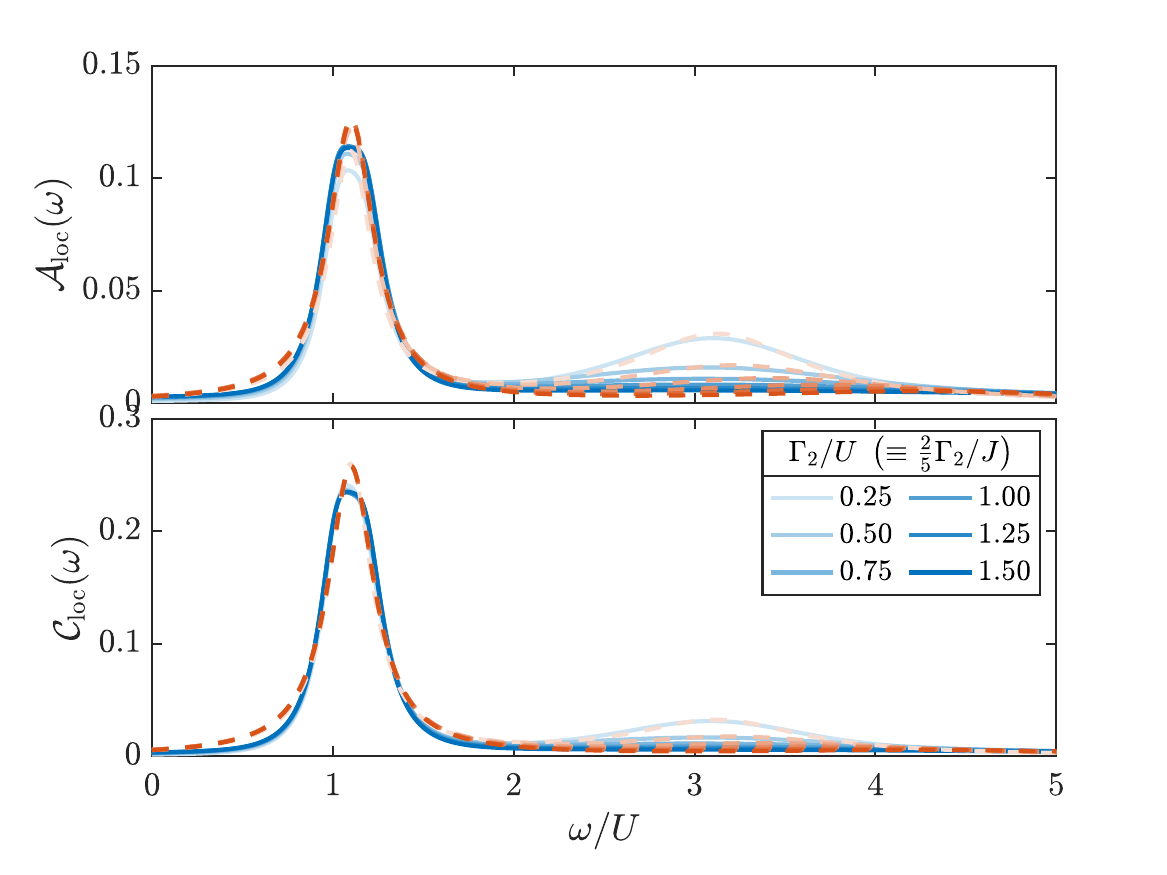}}
    \caption{Local spectral function $\mathcal{A}_{\mathrm{loc}}$ (top panel) and cavity correlation function $\mathcal{C}_{\mathrm{loc}}$ (bottom panel), shown in solid lines for increasing values of $\Gamma_2/U$ (fading blue to strong blue). The red dashed lines correspond to $-z\mathrm{Im}\lbrace\Delta^R\rbrace/(\pi J^2)$ (top panel) and $-z\Delta^K/(2\pi i J^2)$ (bottom panel), and are shaded in accordance with the respective solid blue lines. According to the self-consistency condition on the Bethe lattice (\ref{eq:Bethe_SCF}), at a given value of $\Gamma_2/U$ the solid blue line and its corresponding dashed red line should overlap for an ideal AIM. The color scale and the simulation parameters are as in Fig.~\ref{fig:zeno_emergence_pSS}.}
    \label{fig:zeno_emergence}
\end{figure}

We now consider the local spectral and cavity correlation functions, respectively defined as
\begin{equation}
    \begin{aligned}
        \mathcal{A}_{\mathrm{loc}}(\omega) 
        &= - \frac{1}{\pi} \Imag G^R_{\mathrm{loc}}(\omega) \\
        \mathcal{C}_{\mathrm{loc}}(\omega)
        &= -\frac{1}{2\pi i} G^K_{\mathrm{loc}}(\omega)
    \end{aligned}
    \label{eq:A_C_omega}
\end{equation}
and which encode, respectively, the spectrum and occupation of single-particle excitations on top of the stationary state~\cite{Scarlatella2018,Secli2021}. Once the DMFT self-consistency has been reached, these functions coincide with the analogous quantities at the impurity site; we plot them in Fig.~\ref{fig:zeno_emergence} (solid blue lines), alongside their fitted bath hybridization functions (dashed red lines), for a fixed $J/U=0.4$. At the lowest value of $\Gamma_2/U$ shown in the plot, we still have two visible peaks in the spectral function (top panel), corresponding to the $\ket{0} \to \ket{1}$ transition at low energies and to the $\ket{1} \to \ket{2}$ transition at higher energies. The bath sites, for their part, contribute each one a Lorentzian in the bath hybridization function $\mat{\Delta}$; the Lorentzian contributed by the site $i=1$ fits the peak at low energies ($\omega_1 \approx \omega_0+U$), while the one contributed by the site $i=2$ fits the peak at high energies ($\omega_2 \approx \omega_0+3U$). As $\Gamma_2/U$ is increased, the peak at high energies quickly becomes less and less prominent until it becomes irrelevant in the spectral function; on the other hand, the first peak in the spectral function behaves with the opposite trend. As we comment more in detail in Sec.~\ref{sec:effectivedimer}, this is a signature of the effective narrowing of the Hilbert space to the two Fock states $\ket{0}$ and $\ket{1}$. While the disappearance of the secondary peak is also visible in the cavity correlation function (bottom panel), connected with the occupation, a corresponding increase or decrease in the first peak is not immediately apparent.

\begin{figure}[tbp]
    \centering
    \includegraphics[scale=0.45]{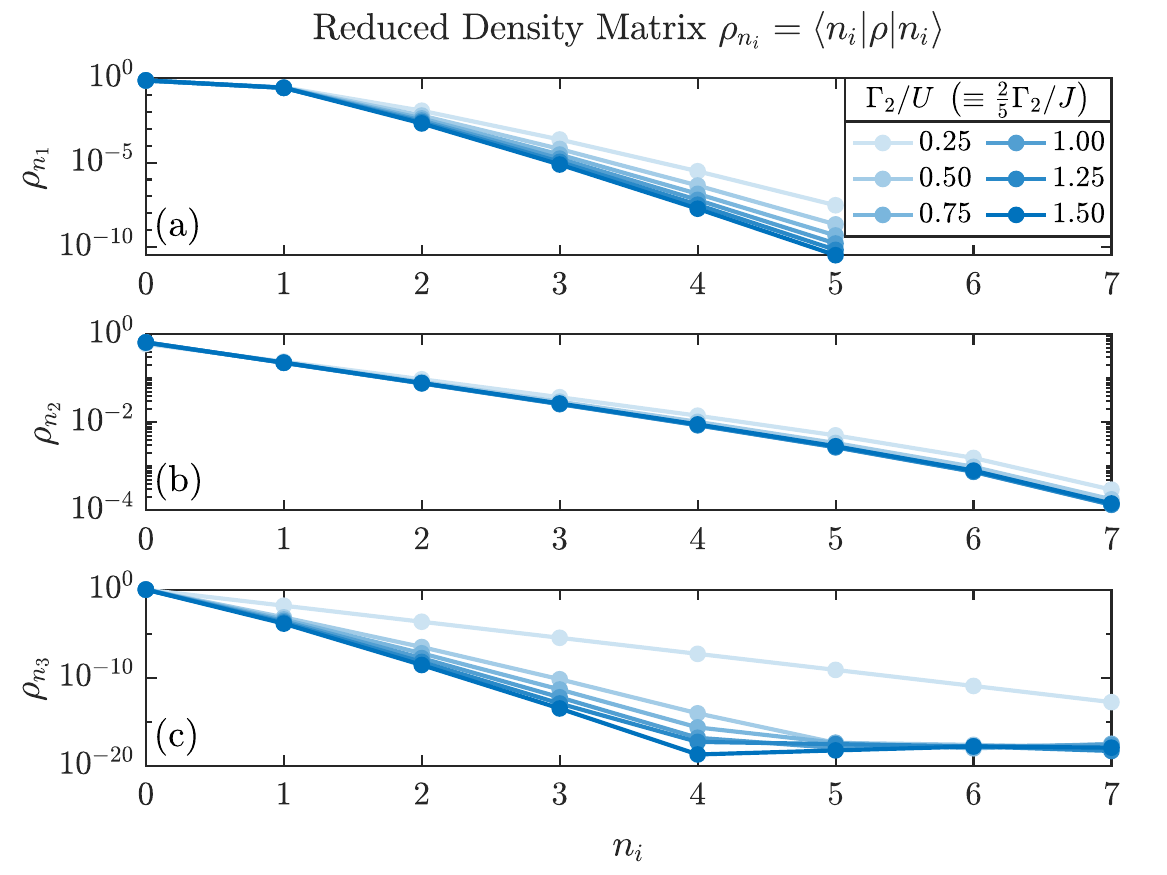}
    \caption{Elements of the AIM on-site reduced density matrix $\rho_{n_i}$ in the Fock basis at site $i$ with $n_i$ bosons. The impurity site is indexed by $i=0$ (panel (a)), while the bath site indexed by $i=1$ ($i=2$) is the one at low (high) energies (panels (b) and (c), respectively). $\Gamma_2/U$ (or equivalently $\Gamma_2/J$) is increased from fading blue to strong blue. Parameters: $z=6$ on the Bethe lattice, $J/U = 0.4$, $P_1/U = 0.01$, $\Gamma_2/U = [0.25,1.50]$. The cutoffs on the $i=0,1,2$ site are respectively $5,7,7$, as commented in the text.}
    \label{fig:zeno_emergence_pSS}
\end{figure}

For the regime considered in this work, the spectral peaks at even higher energies, corresponding to transitions to states with $n>2$, are thus not relevant. Our effective AIM sketched in Fig.~\ref{fig:aim_zeno}, however, provides a great flexibility; in those regimes in which the description of higher excitations is instead important, the AIM can be easily modified by adding further bath sites, with the caveat of an increased dimension of the Hilbert space.

\subsection{QZE, DMFT Bath, and Effective Dimer Model}
\label{sec:effectivedimer}

In Sec.~\ref{sec:steadystatezeno} we have shown that our DMFT/ED approach is able to capture the steady-state QZE in the on-site occupation, a result that is typically beyond single-site mean-field theories like Gutzwiller's. In order to further highlight the origin of this result within DMFT, it is useful to study more in detail the properties of the DMFT bath in the QZ regime.

We plot in Fig.~\ref{fig:zeno_emergence_pSS} the elements of the on-site reduced density matrix in all the three sites of the auxiliary AIM, for a fixed $J/U=0.4$. These elements give the occupation probability in each on-site Fock state. We immediately notice that, on the impurity site (panel (a)), the contributions to the density matrix come mainly from the Fock states $\ket{0}$ and $\ket{1}$; together, these two states contribute (increasing from low to high $\Gamma_2/U$) $98.8\%$--$99.8\%$ of the total weight on the impurity site, thus validating our expectations on the extent of the reduced Hilbert space and our cutoff choice for $i=0$. The situation is quite different in the auxiliary bath sites (panels (b) and (c)), that instead display a thermal-like occupation. The fact that the Fock states $\ket{n}$ with $n \geq 2$ are still relevant even in the deep Zeno regime is the reason we are using a higher cutoff of $7$ in the bath sites with respect to a cutoff of $5$ that we use in the impurity site. An interesting result is the remarkable difference in the order of magnitude of  the occupation probabilities of the two bath sites. In particular, the $i=2$ bath site, at higher energies, becomes less and less relevant as the two-particle losses are increased.  This suggests that deep in the QZ regime, an effective BH dimer model can emerge.

In order to understand the origin of this result, we recall that in the deep Zeno regime ($\Gamma_2 \gg U > J$) we expect that the Fock space on the impurity site is effectively reduced to $\ket{0}$ and $\ket{1}$, since states with $2$ or more particles are strongly dissipated by the two-particle losses. As a consequence, the spectral function on the impurity site has a single peak around $\omega_0 + U$, corresponding to transitions between $\ket{0}$ and $\ket{1}$. This is indeed confirmed if we look at the self-consistent Green's function in Fig.~\ref{fig:zeno_emergence}. However, as also made clear by the Bethe lattice self-consistency condition (\ref{eq:Bethe_SCF}), the auxiliary bath in some sense mirrors the properties of the impurity site (and in turn, those of the lattice model). This means that, since the occupation on the impurity site is low due to the presence of two-particle losses that dissipate states with higher occupations, we expect the occupation on the auxiliary bath sites to be low as well. Specifically, for the parameters discussed here, the occupation on the $i=1$ bath site (panel (b)), having energy $\omega_1 \approx \omega_0+U$, is roughly in the range $0.5$--$0.7$, so based on considerations on the cutoff choice put forward in our previous work~\cite{Secli2021}, a cutoff of $7$ is expected to provide results with a negligible error. This is especially clear for the $i=2$ bath site (bottom panel), at energy $\omega_2 \approx \omega_0+3U$, whose occupation visibly decreases as $\Gamma_2$ is increased, due to the fact that in the deep Zeno regime we expect not to have any other spectral peaks except for the first one, as we indeed observe in Fig.~\ref{fig:zeno_emergence}. At the higher values of $\Gamma_2$ shown here, the occupation probability of Fock states $\ket{n}$ with $n \gtrsim 4$ is already in the order of magnitude of the numerical noise. The sizable difference on the order of magnitude of the occupation probabilities of the two bath sites already points to the key result that the $i=2$ bath site, at higher energies, becomes less and less relevant as the two-particle losses are increased. This effectively allows us to solve the problem, in the deep Zeno regime, via a linearized DMFT, i.e.\@ a DMFT with a single bath site. Interestingly, a similar approach has been proposed as an approximate solution of DMFT in Ref.~\cite{Potthoff2001}.

\begin{figure}[tbp]
    \centering
    \mbox{\hspace*{-12pt}\includegraphics[scale=0.5]{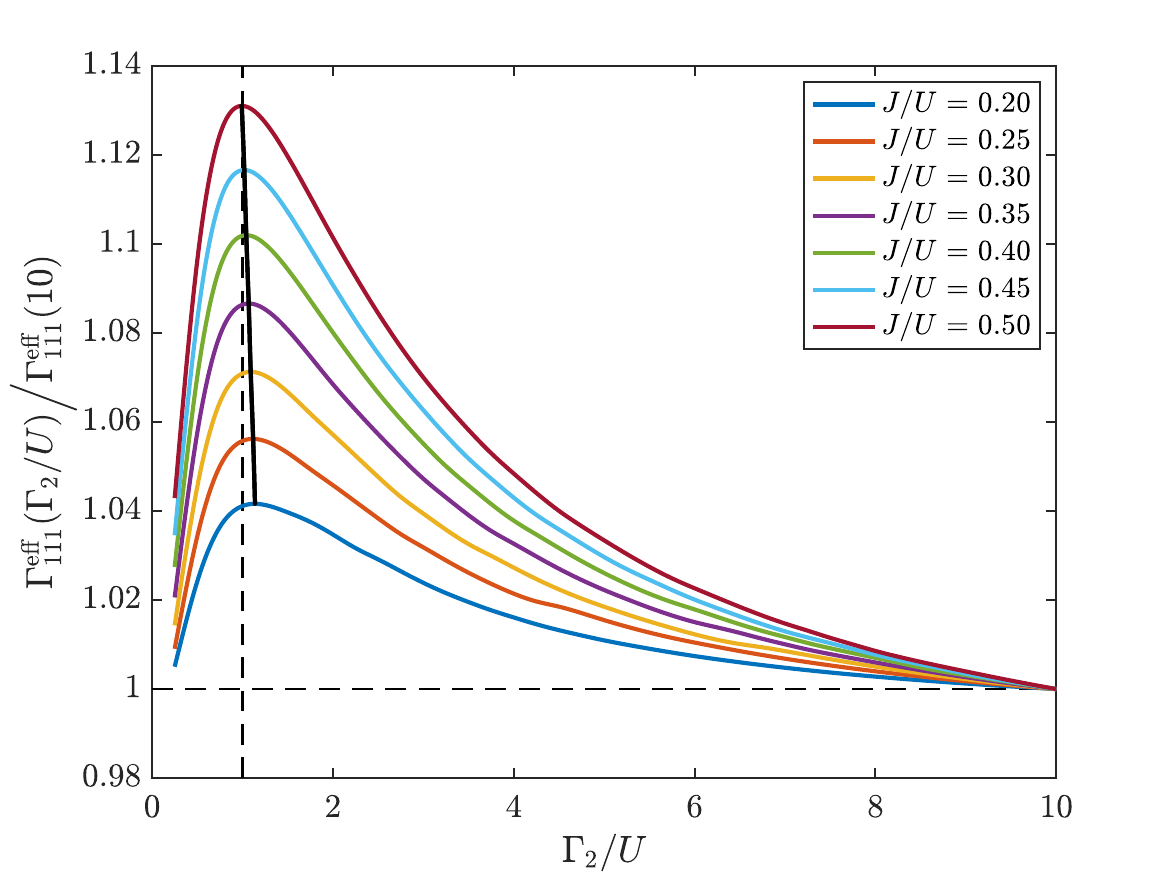}}
    \caption{Effective loss rate $\Gamma_{111}^{\mathrm{eff}} = \Gamma_{111} - P_{111}$ on the first bath site $i=1$ (see also Figs.~\ref{fig:zeno_emergence} and~\ref{fig:zeno_emergence_pSS}), normalized to its value at $\Gamma_2/U = 10$; different colors correspond to different values of $J/U$. The horizontal and vertical dashed black lines mark the values $\Gamma_{111}^{\mathrm{eff}}(\Gamma_2/U)/\Gamma_{111}^{\mathrm{eff}}(10) = 1$ and $\Gamma_2/U = 1$, respectively; the solid black line marks the maximum $\Gamma_{111}^{\mathrm{eff}}$ for different values of $J/U$. Parameters: as in Fig.~\ref{fig:zeno_emergence_pSS}, except for $J=[0.2,0.5]$ and $\Gamma_2 = [0.25,10.0]$.}
    \label{fig:zeno_bathparameters}
\end{figure}

Finally, it is interesting to look for signatures of such a transition in the properties of the auxiliary bath as well, since we already pointed out that the bath itself acts as a mirror image of the original lattice. Since the first bath site ($i=1$) is the only relevant one in the deep Zeno regime, we can analyze its effective (in the semiclassical sense) dissipation rate, given by the difference $\Gamma_{111}^{\mathrm{eff}} = \Gamma_{111} - P_{111}$ between the single-particle dissipation rate and the single-particle pump rate, as a function of $\Gamma_2/U$. This quantity corresponds to the HWHM of the Lorentzian contributed by this site to the bath hybridization function. We see in Fig.~\ref{fig:zeno_bathparameters} that this effective loss rate qualitatively reproduces the behavior of the two-particle effective loss rate $\Gamma_2^{\mathrm{eff}}$, i.e.\@ it first increases up to a maximum value at $\Gamma_2/U \approx 1$, after which it starts decreasing as $\Gamma_2/U$ is increased. The behavior of the effective loss on the first auxiliary bath site provides a remarkably clear picture of the process through which DMFT successfully capture the quantum Zeno effect, through an effective hard-core Bose-Hubbard dimer. In this effective model one site describes the impurity (and thus a given site of the original lattice) and it is therefore dissipationless, being restricted to the two-state manifold by the strong losses, while the other mimics the rest of the lattice and is exposed to single-particle losses, with a rate controlled by the Zeno scale and single-particle pump. Those two mechanisms, together with the impurity-bath coupling, set the local occupation on the impurity site and provide the physical origin for the QZ behavior discussed in Sec.~\ref{sec:steadystatezeno}.

\section{Conclusions}
\label{sec:conclusions}

In this work, we have analyzed a driven-dissipative Bose--Hubbard lattice in the thermodynamic limit, where the driving is achieved via a single-particle incoherent pump at a rate $P_1$ and the local dissipation removes pairs of particles from the system at a rate $\Gamma_2$. In order to perform a quantum treatment despite the sheer size of the Hilbert space, we have employed a formulation of the DMFT technique --- originally developed to study strongly correlated electronic systems --- which deals with driven-dissipative bosonic systems. The core idea of this technique is to replace all the interconnected nonlinear lattice sites surrounding any given lattice site with an effective linear bath, which in our case is in turn parameterized in terms of a finite number of linear auxiliary sites. The parameters of the auxiliary sites are then self-consistently determined in order to correctly reproduce the frequency-dependent local physics of the lattice.

Such a driven-dissipative Bose--Hubbard lattice becomes especially interesting in the presence of strong --- compared to the tunneling rate --- local two-particle losses. In fact, in this limit, the lattice can be mapped into a lattice of hard-core bosons with non-local two-particle losses at a rate $\Gamma_2^{\mathrm{eff}}$ induced by the tunneling rate. Interestingly, when the local two-particle losses $\Gamma_2$ overcome the local Kerr nonlinearity $U$, the non-local two-particle losses $\Gamma_2^{\mathrm{eff}}$ actually start decreasing as $\Gamma_2$ is increased, with a behavior $\Gamma_2^{\mathrm{eff}} \sim \Gamma_2^{-1}$ in the limit $\Gamma_2 \gg U$. This regime is known as the quantum Zeno regime.

In our work we have studied the emergence of a quantum Zeno regime in the stationary state, adding a small single-particle pump to the two-body losses. The steady-state quantum Zeno effect is then testified by the behavior of the number of particles as a function of $\Gamma_2/U$ on a given lattice site. While for $\Gamma_2/U \lesssim 1$ the occupation decreases as the local two-particle losses increase, for $\Gamma_2/U \gtrsim 1$ the quantum Zeno effect results in an increase of such occupation as $\Gamma_2$ is increased. This effect becomes stronger at higher values of $J/U$, and our observations are in qualitative agreement with DMFT results obtained with a NCA impurity solver \cite{Scarlatella2020}.

Using our exact-diagonalization Lindblad impurity solver, we have further highlighted the mechanism by which DMFT is able to capture the Zeno regime, which is not captured by the Gutzwiller mean field theory. We have shown that the structure of the impurity model simplifies in the strongly dissipative limit, where the occupation on the impurity site is restricted to the $\ket{0} \to \ket{1}$ manifold and only one bath site remains effectively populated. Furthermore, the effective dissipation of this impurity bath site is controlled by the Zeno scale. Dynamical mean-field theory provides, therefore, a simple physical picture for the emergence of the steady state Zeno regime in terms of an effective hard-core Bose-Hubbard dimer.

In this work we have implemented~\cite{Secli2022} a straightforward full diagonalization which severely limits the size of the matrices. We can foresee an extension to Arnoldi diagonalization in order to deal with larger Hilbert spaces which are necessary to treat, e.g., systems with a finite condensate fraction and with coherent pumping mechanisms. This will allow exploring systems with light-matter coupling at the quantum level \cite{Biella2017,Lebreuilly2017,Lebreuilly2017a,Ma2019} and with non-trivial non-Markovian effects, making our OpenBDMFT implementation a promising platform for the investigation of novel many-body physics.

\acknowledgments

M. Capone and M.Seclì acknowledge financial support by the Italian MIUR under the PRIN2017 project CEnTral (Protocol Number 20172H2SC4). M. Schirò acknowledges support by the ANR grant "NonEQuMat" (ANR-19-CE47-0001).
M. Seclì acknowledges the CINECA award HP10CFGJ44 (2021) under the ISCRA initiative, for the availability of high performance computing resources and support, as well as SISSA, for the availability of high-performance computing resources on the Ulysses cluster.
M. Seclì would also like to thank Fabio Caleffi for continuous and stimulating discussions.

\appendix

\section{AIM Diagonalization}
\label{app:aimdiag}

In this Appendix we provide additional details on the vectorization of the Lindblad equation and on the exact-diagonalization impurity solver we have developed. For a more in-depth discussion on our specific implementation, we refer the interested reader to Refs.~\cite{Secli2021b,Secli2022}.

The AIM in Eqs.~(\ref{eq:AIM_Hermitian_evolution}) to~(\ref{eq:AIM_Lindblad_dissipator}) can be numerically solved by diagonalizing a matrix representing the action of the Lindbladian in an enlarged Fock space. In such an enlarged Fock space, the density matrix is represented as a vector, hence this mapping is commonly called \emph{vectorization}.

A generic state of the enlarged Fock space is written as $\ket{\vec{n};\vec{\tilde{m}}} \equiv \ket{\vec{n}} \otimes \ket{\vec{\tilde{m}}}$, where $\ket{\vec{n}} \equiv \ket{n_0} \otimes \ket{n_1} \otimes \ldots \otimes \ket{n_{N_B}}$ is a Fock state of the original AIM, while $\ket{\vec{\tilde{m}}} \equiv \ket{\tilde{m}_0} \otimes \ket{\tilde{m}_1} \otimes \ldots \otimes \ket{\tilde{m}_{N_B}}$ is a Fock state of a duplicate of the original AIM, marked by an additional tilde ``$\sim$''. In this notation, the subscript ``$0$'' indicates the impurity site, i.e. $\hat{n}_0 = \hat{a}^{\dagger}\hat{a}$ and, for $j>0$, $\hat{n}_j = \hat{b}_j^{\dagger}\hat{b}$. Particles in the tilde-copy of the system are created by bosonic operators $\left[\hat{\tilde{a}},\hat{\tilde{a}}^{\dagger}\right] = 1$ and $\left[\hat{\tilde{b}}_i,\hat{\tilde{b}}_j^{\dagger}\right] = \delta_{ij}$, which always commute with their respective non-tilde operators \cite{Ojima1981}.

In this enlarged Fock basis, the vector representing the density matrix is obtained as $\ket{\rho} = \hat{\rho} \ket{I}$, where
\begin{equation}
    \ket{I} = \sum_{\vec{n}} \ket{\vec{n};\vec{\tilde{n}}}\,.
\end{equation}
Once the Lindbladian is written as a matrix $\hat{\mathcal{L}}$, $\ket{I}$ has, by construction, the key property that $\bra{I}\hat{\mathcal{L}} = 0$, hence it's called the \emph{left vacuum}. In the vectorized representation, the trace of a generic operator $\hat{\mathcal{O}}$ is obtained as $\Tr\left(\hat{\mathcal{O}}\right) = \braket{I|\hat{\mathcal{O}}|I}$, while its expectation value is given by $\braket{\hat{\mathcal{O}}} = \Tr\left(\hat{\mathcal{O}}\hat{\rho}\right) = \braket{I|\hat{\mathcal{O}}|\rho}$. In particular, we obtain that $\braket{I|\rho} = 1$, which is the vectorized equivalent of the requirement that the density matrix has a unity trace: $\Tr\rho = 1$. In the following, we will drop the operator hats for notational convenience.

When acting on the left vacuum $\ket{I}$, an operator $A$ acting on the original system can be exchanged for its respective $\widetilde{A}$ acting on the tilde-system according to the following \emph{tilde-conjugation rule}:
\begin{equation}
    A \ket{I} = \sigma_A\widetilde{A}^{\dagger}\ket{I},
    \label{eq:tilde_conj_rule}
\end{equation}
where
\begin{equation}
    \sigma_A =
    \begin{cases}
        -i & \text{if $A$ is a fermionic operator} \\
        1 & \text{if $A$ is a bosonic operator}
    \end{cases}
    \label{eq:tilde_conj_rules_general}
\end{equation}

Thanks to this property, the Lindblad equation transforms into the vectorized form
\begin{equation}
    \frac{d}{dt}\ket{\rho} 
    = \mathcal{L} \ket{\rho}
    = \mathcal{L}_H \ket{\rho} + \mathcal{L}_D \ket{\rho},
\end{equation}
where, for a system described by a Hamiltonian $H$ and by a generic dissipator
\begin{equation}
    \mathcal{L}_D \rho
    = \sum_{\alpha} 2\gamma_{\alpha} \left( L_{\alpha} \rho L_{\alpha}^{\dagger} - \frac{1}{2}\left\lbrace L_{\alpha}^{\dagger}L_{\alpha},\rho \right\rbrace \right)\,,
\end{equation}
the vectorized representation yields
\begin{equation}
    \mathcal{L}_{H}\ket{\rho}
    = -i\Big( H - \widetilde{H} \Big) \ket{\rho}
\end{equation}
and
\begin{equation}
    \mathcal{L}_{D}\ket{\rho}
    = \sum_{\alpha} \gamma_{\alpha}\left( \sigma_{L_{\alpha}} L_{\alpha}\tilde{L}_{\alpha} - L_{\alpha}^{\dagger}L_{\alpha} - \tilde{L}_{\alpha}^{\dagger}\tilde{L}_{\alpha} \right) \ket{\rho} \,.
    \label{eq:tilde_dissipator}
\end{equation}

\begin{figure}
    \centering
    \includegraphics[scale=0.5]{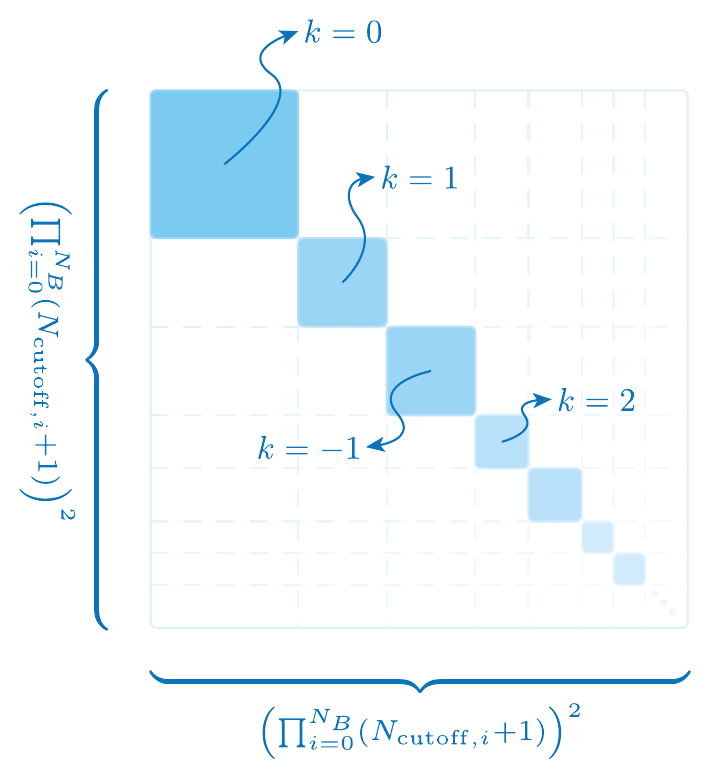}
    \caption{Block-diagonal structure for the gauge-symmetric Lindbladian of the AIM discussed in Sec.~\ref{sec:lindblad_dmft_ed}, with $N_B$ bath sites plus the impurity, the latter indexed by ``$0$''. The full Lindbladian matrix has a size $\left(\prod_{i=0}^{N_B}(N_{\mathrm{cutoff},i}+1)\right)^2 \times \left(\prod_{i=0}^{N_B}(N_{\mathrm{cutoff},i}+1)\right)^2$, but it can be written as a block-diagonal matrix where each block is labeled by an integer $k$, with $ik$ an eigenvalue of $\doublehat{\mathcal{K}} \bullet = -i \left[ \hat{N}, \bullet \right]$.}
    \label{fig:block_lindbladian}
\end{figure}

Note that the Lindbladian employed in Sec.~\ref{sec:lindblad_dmft_ed} possesses a global $U(1)$ gauge symmetry, i.e.\@ all the particle creation/annihilation operators can be rotated by a global phase.

This symmetry is expressed by the presence of a superoperator $\doublehat{\mathcal{K}}$ that commutes with $\doublehat{\mathcal{L}}$, which is simply the superoperator generated by the total number of particles, i.e.\@ $\doublehat{\mathcal{K}} \bullet = -i \left[ \hat{N}, \bullet \right]$, where $\hat{N} = \sum_{i=0}^{N_B} \hat{n}_i$.

In the vectorized representation, $\doublehat{\mathcal{K}}$ assumes a particularly simple form: $\mathcal{K} = i\Big( \tilde{N} - N \Big) = ik$, where, since we work in the number basis, the linear operator $\mathcal{K}$ has eigenvalues $ik$, $k = 0,\,+1,\,-1,\,+2,\,-2,\,\ldots$.

Since the Lindbladian and $\mathcal{K}$ commute, they share a common set of eigenvectors; hence we can classify the eigenvectors of the Lindbladian by labeling them with the eigenvalues of $\mathcal{K}$, i.e.\@ with $k$ if we drop the imaginary unit for convenience. In other words, the Lindbladian can be written as a block-diagonal matrix, with each block (also called ``sector'') labeled by an integer value $k$ which is equal to the difference between the occupation in the tilde-system and the occupation in the original system --- see Fig.~\ref{fig:block_lindbladian}.

Our \texttt{C++} OpenBDMFT code \cite{Secli2022} can perform the numerical diagonalization via the standard \texttt{LAPACK} (or its high-speed replacements) libraries thanks to a user-friendly interface provided by the \texttt{Armadillo} library \cite{Sanderson2016,Sanderson2018}, that we have extended to support the two-sided diagonalization of non-symmetric matrices. However, due to she sheer size of the problem, the diagonalization times via CPU-based routines like \texttt{LAPACK}'s ones are particularly high. In order to further cut down on the diagonalization times, we eventually customized \texttt{Armadillo}'s internal diagonalization routines so to accelerate the diagonalization via GPUs, when they are available. The support for GPU acceleration is provided via the \texttt{MAGMA} library~\cite{Tomov2010,Tomov2010a,Dongarra2014}.

The diagonalization process yields the eigenvalues $\mathcal{L}_{\alpha}$ as well as the bi-normalized left/right eigenvectors $\bra{l_{\alpha}}$ and $\ket{r_{\alpha}}$ of the Lindbladian $\hat{\mathcal{L}}$:
\begin{equation}
    \bra{l_{\alpha}} \hat{\mathcal{L}} = \mathcal{L}_{\alpha} \bra{l_{\alpha}}
    \qquad\text{and}\qquad
    \hat{\mathcal{L}} \ket{r_{\alpha}} = \mathcal{L}_{\alpha} \ket{r_{\alpha}}
\end{equation}
where the bi-orthonormalization ensures that $\braket{l_{\alpha}|r_{\beta}} = 1$.

This spectral information can be used to numerically calculate the single-particle Green's function on the impurity site which, as discussed in Sec.~\ref{sec:dmft_loop}, is the central object of any DMFT calculation. In particular, the retarded/Keldysh components of the impurity Green's function in Eqs.~(\ref{eq:G_imp_R_defs}) and~(\ref{eq:G_imp_K_defs}) have the following spectral representation \cite{Secli2021}:
\begin{align}
    G_{\mathrm{imp}}^R(\omega)
    &= \sum_{\alpha}\Braket{I|a|r_{\alpha}}\Braket{l_{\alpha}|a^{\dagger}|\rho_{\mathrm{ss}}}\frac{1}{\omega - i\mathcal{L}_{\alpha}} \nonumber \\
    &- \left(\sum_{\alpha}\Braket{I|a^{\dagger}|r_{\alpha}}\Braket{l_{\alpha}|a|\rho_{\mathrm{ss}}}\frac{1}{\omega + i\mathcal{L}_{\alpha}}\right)^*
    \label{eq:G_R_KL_W}
\end{align}
\begin{align}
    G_{\mathrm{imp}}^K(\omega)
    &= 2 i\Imag \left( \sum_{\alpha}\Braket{I|a|r_{\alpha}}\Braket{l_{\alpha}|a^{\dagger}|\rho_{\mathrm{ss}}}\frac{1}{\omega - i\mathcal{L}_{\alpha}} \right) \nonumber \\
    &+ 2 i\Imag \left( \sum_{\alpha}\Braket{I|a^{\dagger}|r_{\alpha}}\Braket{l_{\alpha}|a|\rho_{\mathrm{ss}}}\frac{1}{\omega + i\mathcal{L}_{\alpha}} \right)^*
    \label{eq:G_K_KL_W}
\end{align}

\vbox{}                         

\bibliographystyle{apsrev4-1}   
\typeout{}                      
\bibliography{references}

\end{document}